\documentclass[prb,aps,twocolumn]{revtex4}
\usepackage{graphicx,color}
\begin{document}

\title{Phase separation in the vicinity of the surface of 
$\kappa$-(BEDT-TTF)$_2$Cu[N(CN)$_2$]Br by fast cooling} 
\author{N. Yoneyama, T. Sasaki, and N. Kobayashi}
\address{Institute for Materials Research, Tohoku University, Sendai
980-8577, Japan}
\author{Y. Ikemoto and H. Kimura}
\address{Spring-8, Japan Synchrotron Radiation Research Institute, Mikazuki, 
Hyogo 679-5198, Japan}

\date{\today}

\begin{abstract}

Partial suppression of superconductivity by fast cooling has 
been observed in the organic superconductor
$\kappa$-(BEDT-TTF)$_2$Cu[N(CN)$_2$]Br by two means:
a marked sample size effect on the magnetic susceptibility and direct
imaging of insulating regions by scanning microregion infrared reflectance spectroscopy.
Macroscopic insulating regions are proposed in the vicinity of the 
crystalline surface after fast cooling, with diameters of 50--100 $\mu$m and depths of a few $\mu$m.
The very large in-plane penetration depth reported to date ($\sim$ 
24--100 $\mu$m) can be explained by the existence of the insulating 
regions. 

\end{abstract}
\pacs{74.70.Kn, 71.30.+h}

\maketitle

\section{Introduction}
\begin{figure}
\includegraphics[width=6.5cm,clip]{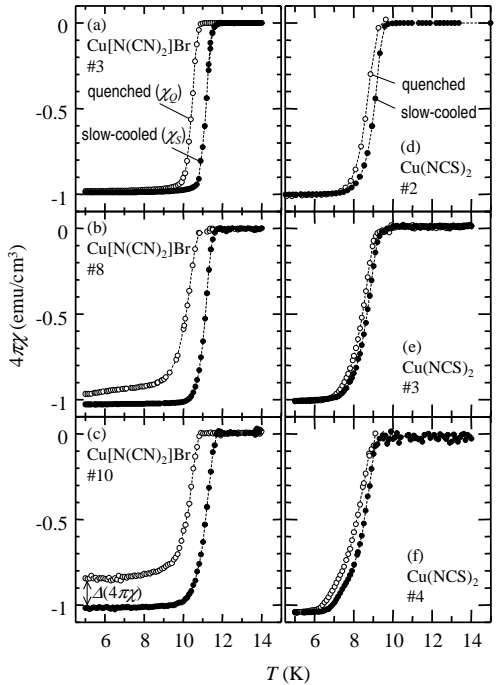}
 \caption{\label{Fig:chiT}
Temperature dependence of the magnetic susceptibility for 
$\kappa$-(h8ET)$_2X$, where $X$=Cu[N(CN)$_2$]Br for panels (a)--(c) and 
$X$=Cu(NCS)$_2$ for panels (d)--(f). Samples with larger index numbers have smaller 
crystalline sizes.\cite{size_Br,size_CuNCS}
 }
\end{figure}

The magnetic penetration depth $\lambda$ is one of the most 
fundamental parameters of superconductivity.
It reflects the excitation of quasiparticles, governed
by the superconducting (SC) gap symmetry.
To clarify the mechanism of superconductivity,
a careful study of the temperature dependence of $\lambda$ is 
essential.

There have been many previous studies on the in-plane penetration depth 
$\lambda_{\parallel}$ of the quasi-two-dimensional organic superconductors 
$\kappa$-(BEDT-TTF)$_2X$ [$X$=Cu(NCS)$_2$ or Cu[N(CN)$_2$]Br; BEDT-TTF: 
bis(ethylenedithio)-tetrathiafulvalene, also written as ET],
by ac 
susceptibility,\cite{793,153,20,785} surface impedance,\cite{994,141} 
ac inductance\cite{108}, $\mu$SR,\cite{995,177,163,181} 
magnetization,\cite{197,180,220,1151} etc.
In spite of these efforts, the temperature dependence of 
$\lambda_{\parallel}$ and the absolute value of $\lambda_{\parallel}(0)$ 
are still unknown or controversial.
A precise determination of $\lambda_{\parallel}(0)$ is 
important for the superfluid density  
$\rho(T)=[\lambda_{\parallel}(T)/\lambda_{\parallel}(0)]^{-2}$; 
the behavior of $\rho(T)$
is strongly dependent on $\lambda_{\parallel}(0)$.\cite{108}
However, the reported values of $\lambda_{\parallel}(0)$ for the two superconductors have large 
distributions, 0.4--2 $\mu$m [$X$=Cu(NCS)$_2$] and 0.57--100 $\mu$m 
[$X$=Cu[N(CN)$_2$]Br].\cite{1151}
We have recently found that these values of $\lambda_{\parallel}(0)$ can be 
classified into two groups:\cite{1151} one for 
$\lambda_{\parallel}(0)$ larger than 1 $\mu$m and the other for shorter $\lambda_{\parallel}(0)$.
The data in the first group were measured in magnetic fields lower 
than $H_{c1}$, in which a shielding current flows at the sample 
surface (shielding state).
In this case, $\lambda$ indicates the length from the surface $H$ penetrates the sample.
The data in the second group were measured in magnetic 
fields higher than $H_{c1}$, in which $H$ penetrates into the sample as 
vortices (mixed state).
In this case, $\lambda$ is a decay length of the magnetic field 
from the center of a vortex.

A remarkable difference in $\lambda_{\parallel}(0)$ between the two 
groups 
appears for fast cooling, especially for $X$ = Cu[N(CN)$_2$]Br.
In ac susceptibility measurements\cite{785} in magnetic fields lower than 
$H_{c1}$, $\lambda_{\parallel}(0)$ is 24--100 $\mu$m, which is two or 
three orders of magnitude larger than that obtained from magnetization 
measurements\cite{220,1151} in magnetic fields higher than $H_{c1}$, 
$\sim$ 0.7 $\mu$m.
The former value of $\lambda_{\parallel}(0)$ is quite large as a characteristic 
parameter of superconductivity and is likely to be an overestimate.
This measurement may be sensitive to the conditions
of the first several layers of the surface, e.g., edge, step, and 
surface reconstruction.\cite{310}
However, the fast cooling data of $\lambda_{\parallel}(0)$ in the 
low field measurements are still too large to be explained by 
these surface effects in the nanometer scale.
This implies the existence of non-SC regions in the vicinity of the 
surface, probably in the macroscopic scale. 

In the present report, we suggest that the SC state in
$\kappa$-(h8ET)$_2$Cu[N(CN)$_2$]Br (abbreviated as $\kappa$-h8ET-Br, where h8ET is conventional 
hydrogenated BEDT-TTF) 
after fast cooling coexists with non-SC (insulating) portions in the micrometer scale in 
the vicinity of the crystalline surface.
We prove this by two experiments.
In the first experiment, a marked size effect on the SC volume fraction is obtained from 
static magnetic susceptibility measurements.
On the basis of a crude model, we propose that the mean depth of 
the insulating region is a few $\mu$m from the surface.
In the second experiment we directly image an insulating region of 
50--100 $\mu$m diameter by scanning microregion infrared 
reflectance spectroscopy (SMIS) measurements.\cite{1208,1308}
This technique is insensitive to the surface condition of the first 
several layers and thus is suitable for the present study.
The very large reported $\lambda_{\parallel}(0)$ ($\sim$ 24--100 
$\mu$m) can be reasonably well explained by the existence of thin insulating
domains in the surface vicinity.

\section {Experiments}
\begin{figure}
\includegraphics[width=6.5cm,clip]{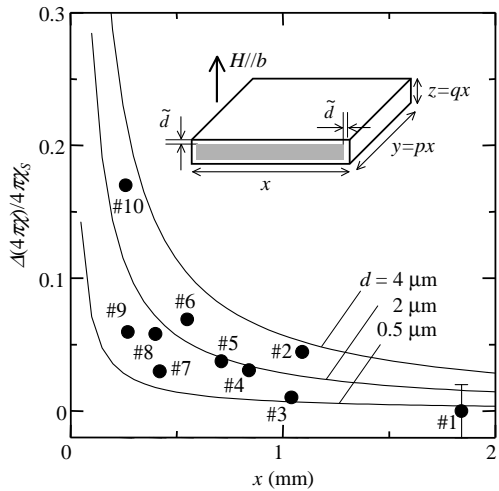}
 \caption{\label{Fig:Vx}
Suppression of superconducting volume fraction by quenching 
in $\kappa$-(h8ET)$_2$Cu[N(CN)$_2$]Br.
The inset shows a schematic view of the sample dimensions.
Based on a rectangular shell-type insulating model, the solid curves 
are obtained from Eq. \ref{Eq:Vx} for shell thicknesses of
$\tilde{d}$ = 0.5, 2, and 4 $\mu$m.
}
\end{figure}

Single crystals of $\kappa$-h8ET-Br were grown by a standard electrochemical oxidation technique.
The crystalline shape of $\kappa$-h8ET-Br is typically 
rectangular with (010) rhombus facets (inset of Fig. \ref{Fig:Vx}). 

Static magnetic susceptibility measurements were performed using a SQUID 
magnetometer (Quantum Design, MPMS-5XL).
To investigate the crystalline size effect, we chose 10 single 
crystals (\#1 to \#10).\cite{size_Br}
The samples with larger index numbers have smaller crystalline size.
A magnetic field of 3 Oe was applied perpendicular to the conduction 
plane.
The susceptibilities were measured in a zero-field cooling condition, 
giving shielding curves.
A very fast cooling procedure ($\sim$100 K/min) from room temperature 
to 15 K was first adopted, giving the ``quenched'' (\textit{Q}) state. 
Then a second cooling process was carried out with 0.2 K/min 
after warming to 100 K, giving the ``slow-cooled'' (\textit{S}) state.
The demagnetization factor was corrected using an ellipsoidal approximation.

SMIS studies were performed using a synchrotron radiation beam 
at the BL43IR in SPring-8.\cite{1209}
Polarized reflectance spectra were measured on the conductive
$ca$-plane with the electric field parallel to the $a$-axis
by use of a Fourier transform infrared (FTIR) spectrophotometer from 
500 to 2500 cm$^{-1}$.
A single crystal of $\kappa$-h8ET-Br with dimensions approximately
$0.5\times0.3\times0.1$ mm$^3$ (the same sample as used in Ref. \onlinecite{1308}) 
was attached to a thermal anchor with carbon paste.
The sample was cooled slowly at a rate of $\sim$ 0.4 K/min from 
room temperature to 4 K, giving the \textit{S} state.
After the SMIS measurement in the \textit{S} state,
the temperature was increased to 120 K and then cooled at a rate of
$\sim$ 35 K/min, giving the \textit{Q} state.

\section{Results}
\begin{figure}
\includegraphics[width=6.5cm,clip]{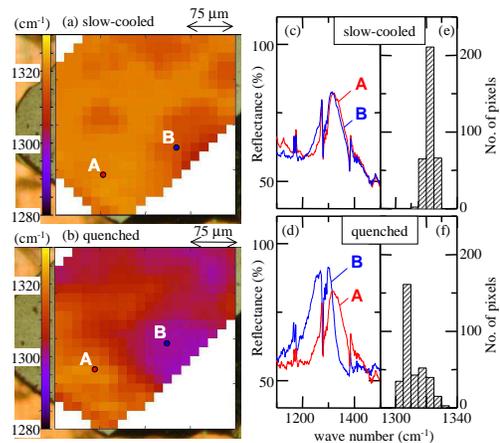}
 \caption{\label{Fig:SMIS}
(color online)
(a)\cite{1308}, (b) Peak frequency maps of the $\nu_3 (a_g)$ mode of 
$\kappa$-(h8ET)$_2$Cu[N(CN)$_2$]Br at 4 K.
(c), (d) Reflectance spectra at A- and B-sites 
in Figs. \ref{Fig:Vx} (a) and (b), respectively. (e), (f) Histograms of the $\nu_3 (a_g)$ peak 
frequency.
}
\end{figure}

Figures \ref{Fig:chiT}(a)--\ref{Fig:chiT}(c) show the temperature dependences of the 
magnetic susceptibility in $\kappa$-h8ET-Br for 
single crystals with different sample sizes.
Regardless of sample size, an almost perfect Meissner effect is 
realized in the \textit{S} state (filled circles).
In the large sample \#3, shown in Fig. \ref{Fig:chiT}(a), there 
is no change in the SC volume between the \textit{S} and \textit{Q}
states within the experimental error.
This is consistent with previous reports.\cite{1151,203}
In contrast, suppression of the SC
volume in the $Q$ state can be seen with decreasing sample size (for samples \#8 and \#10 in 
Figs. \ref{Fig:chiT}(b) and \ref{Fig:chiT}(c), respectively).
The difference in the SC volume between the two cooling 
conditions, $\Delta (4\pi \chi) = 4\pi \chi_{Q}-4\pi \chi_{S}$, is 
pronounced in the smaller crystals, where $4\pi \chi_{Q}$ and $4\pi
\chi_{S}$ are for the \textit{Q} and 
\textit{S} states at 5 K, respectively.
In Fig. \ref{Fig:Vx}, the values of $\Delta (4\pi \chi)/4\pi \chi_{S}$ 
in $\kappa$-h8ET-Br are plotted as a function of $x$, 
representing the fraction of the non-SC portion, where $x$ is the 
longest length of the crystal (see inset in Fig. \ref{Fig:Vx}).
In spite of large scatter in the data, $\Delta (4\pi \chi)/4\pi 
\chi_{S}$ increases with decreasing $x$.
The data scattering is due mainly to the sample dependence of the 
distribution and morphology in the insulating regions.

To confirm whether this anomalous phenomenon is characteristic of
$\kappa$-h8ET-Br, similar measurements were performed
for four samples of $\kappa$-(h8ET)$_2$Cu(NCS)$_2$.\cite{size_CuNCS}
The magnetic susceptibilities are shown in Figs. 
\ref{Fig:chiT}(d)--\ref{Fig:chiT}(f).
Although $T_{c}$ in the \textit{Q} state is slightly lower by 
about 0.2--0.3 K than in the $S$ state, an almost perfect 
Meissner effect is obtained for both the cooling states, independent of 
sample size. 
We therefore conclude that the effect of crystalline size on SC suppression
is a peculiar feature in quenched $\kappa$-h8ET-Br.
As directly indicated below, the non-SC region in the $Q$ state is 
interpreted as a coexisting insulating phase in the surface vicinity.

In SMIS measurements, a shift of the fully symmetric molecular 
vibration mode $\nu_3(a_g)$ can be a relevant indicator of the 
electronic state, because it is very sensitive to the difference 
between the SC (metallic) and insulating phases.\cite{1115}
Peak frequency maps of the $\nu_3$ mode are shown in Figs. 
\ref{Fig:SMIS}(a)\cite{1308} and \ref{Fig:SMIS}(b) for the \textit{S} and \textit{Q}
states, respectively.
In these maps brighter colors represent higher frequencies.
In the \textit{S} state, the peak frequency shows good spatial homogeneity
at approximately 1320 cm$^{-1}$.
The SC phase gives a higher frequency
in the $\nu_3$ mode at low temperatures,\cite{1115}  
demonstrating that the homogeneous SC state 
appears in the \textit{S} state.
On the other hand, in the map of the \textit{Q} state (Fig. 
\ref{Fig:SMIS}(b)), we find dark areas with much 
lower frequency than for the SC phase.
As shown in Fig. \ref{Fig:SMIS}(d), the peak of the $\nu_3$ mode in the dark 
area (B-site) is shifted toward lower frequency compared with that in
the remaining bright area (A-site). 
The corresponding reflectance spectra at almost the same positions 
in the \textit{S} state are very similar (Fig. \ref{Fig:SMIS}(c)).
Thus, the inhomogeneous frequency map in the \textit{Q} state clearly 
indicates the coexistence of SC and insulating regions.
Figures \ref{Fig:SMIS}(e) and \ref{Fig:SMIS}(f) show peak frequency histograms.
The histogram in the \textit{S} state has a narrow maximum at 
1320--1330 cm$^{-1}$, whereas that in the 
\textit{Q} state is spread over 1300--1330 cm$^{-1}$ with 
a double peak structure,
reflecting spatial inhomogeneity in the \textit{Q} state.

\section{Discussion}

In the phase diagram of the $\kappa$-(ET)$_2X$ system,\cite{1106} the SC and 
antiferromagnetic Mott insulating (AFI) phases appear next to one another.
The ground state in $\kappa$-h8ET-Br is the SC phase,
while the fully deuterated salt $\kappa$-(d8ET)$_2$Cu[N(CN)$_2$]Br
(abbreviated as $\kappa$-d8ET-Br, where d8ET is fully-deuterated BEDT-TTF),
has an AFI ground state accompanied by a minor SC phase (phase 
separation).\cite{144,749}
Deuteration therefore changes the ground state from the SC to AFI phase through a 
first-order Mott transition.\cite{203,1210}
This is likely caused by a structural modulation in the 
slight change of the bond length between C-H and C-D bonds,
i.e., chemical pressure.
The decrease of the bandwidth $W$ by deuteration results in
an increase in $U/W$, where $U$ is the on-site Coulomb interaction.
This leads to a crossing of the Mott boundary toward the AFI ground state.

The minor SC phase in $\kappa$-d8ET-Br is further suppressed by faster cooling,
though it was believed that superconductivity in $\kappa$-h8ET-Br is
robust even after fast cooling.\cite{203,818,146,1030,1151,1210} 
Furthermore, in $^{13}$C-NMR studies of $\kappa$-h8ET-Br after fast 
cooling, no trace of the AFI phase has been observed.\cite{KMpc}
On the contrary, a marked crystalline size effect on the SC 
volume fraction in the $Q$ state is found, reflecting macroscopic suppression of 
superconductivity
and an insulating region with 50--100 $\mu$m diameter is directly observed 
by the SMIS technique. 

Recently, macroscopic phase separation has been found near
the Mott transition for the d8ET molecular substituted system 
$\kappa$-[(h8ET)$_{1-x}$(d8ET)$_x$]$_2$Cu[N(CN)$_2$]Br by the same SMIS 
measurement.\cite{1208,1308}
For the $Q$ state with $x$ = 0.7, it was observed that the dominant 
insulating region contains a small SC region.\cite{1308}
This corresponding to a SC volume fraction suppressed to much
less than $\sim$ 10\%.\cite{1308}
Similarly, the present insulating region observed in quenched
$\kappa$-h8ET-Br (an end-member with $x$ = 0) seems to 
grow over a wide area of 50--100 $\mu$m diameter.
At first sight, one might regard the result for $x$ = 0 as evidence that 
large portions of the sample volume become insulating.
This is however inconsistent with the large SC volume
even in the $Q$ state of 80--90\% (Fig. \ref{Fig:chiT}(c)).
We here note that the skin depth $\delta$ in the SMIS measurement,
$\delta = c/(2\pi \omega \sigma )^{1/2}$, is roughly estimated to be 
0.7 $\mu$m, where $c$ is the speed of light and using 
the values $2\pi\omega \simeq$ 1000 cm$^{-1}$ and $\sigma \simeq 1000 $ S/cm.\cite{1115}
Thus, although the SMIS measurements provide satisfactory bulk 
properties at a depth of approximately 1 $\mu$m, they do not give information on 
regions much deeper than $\delta$ from the crystalline surface.
Taking account of absence of the insulating regions detectable in 
NMR\cite{KMpc}, the existence of the insulating domains at inner site
will be extracted. 
Accordingly, we suggest that the insulating regions in quenched
$\kappa$-h8ET-Br are located in the surface vicinity and 
the depth of the thin domain is distributed over the surface.
To verify this proposal, detailed NMR measurements 
for a quenched single crystal with several sample sizes will 
be required.

In the following, we discuss the mean depth scale of the thin domain.
A crude model can explain the SC suppression in the \textit{Q} state, 
suggesting that the insulating region extends down to a depth of $\sim$ 
2 $\mu$m. 

For the magnetic susceptibility measurements, we chose samples with a 
similar dimensional ratio of $x/x$:$y/x$:$z/x$ $\simeq$ 1:0.9:0.2.
We here adopt the longest length $x$ as the parameter representing
the sample size, namely, $y\simeq px$, $z\simeq qx$, and the sample volume 
$V_0 \simeq pqx^3$, where $p$ and $q$ are dimensionless numbers,
$p = 0.9$ and $q = 0.2$.
The first finding is that an almost perfect
Meissner effect is observed in the \textit{S} state, regardless 
of $x$.
Thus the SC volume in the \textit{S} state ($V_{S}^{SC}$) can be defined as the 
crystalline volume: $V_{S}^{SC} \simeq V_0$.
Of course the SC volume is not the same as $V_0$, because
the magnetic field penetrates into the surface by $\lambda_{\parallel}$;
however, it is sufficiently negligible, as shown below.
The second finding is that the SC volume decreases in 
smaller samples by quenching,
i.e., a smaller $x$ gives a larger $\Delta (4\pi\chi)$, as shown in Fig. 
\ref{Fig:Vx}.
To consider $\Delta (4\pi\chi)$ as a function of $x$,
we here assume an SC state with a rectangular insulating shell:
superconductivity in the \textit{Q} state 
is suppressed in a mean depth $\tilde{d}$ from all the surfaces and edges
(i.e., $\tilde{d}$ is the thickness of the thin shell shown in the 
inset of Fig. \ref{Fig:Vx}) and $\tilde{d}$ is independent of 
sample size.
On the basis of this assumption, the SC volume in the \textit{Q} state is 
$V_{Q}^{SC} \simeq (x-2\tilde{d})(y-2\tilde{d})(z-2\tilde{d})$.
We finally obtain the non-SC fraction in the \textit{Q} state as 
\begin{equation}
(V_{S}^{SC}-V_{Q}^{SC})/V_{S}^{SC} \simeq a\tilde{d}/x
\label{Eq:Vx}
\end{equation}
for $\tilde{d} \ll x$, where $a\equiv 2(p+q+pq)/pq = 14.2$.
This fraction can be compared with the experimental values of
$\Delta (4\pi \chi)/4\pi \chi_{S}$.
In Fig. \ref{Fig:Vx}, $(V_{S}^{SC}-V_{Q}^{SC})/V_{S}^{SC}$ is 
depicted as a function of $x$ for $\tilde{d}$ = 0.5, 2, and 4 $\mu$m (solid curves).
One can see that the curve for $\tilde{d}$ = 2 $\mu$m is in good accordance 
with the experimental data.
This value of $\tilde{d}$ is adequate for the present system, because it 
satisfies all the following conditions:
(i) $\tilde{d} \ll x$ ($\simeq$ 200 $\mu$m),
(ii) $\tilde{d} \geq \lambda_{\parallel}(0)$ ($\simeq$ 0.6--0.7 $\mu$m\cite{1151}),
and (iii) $\tilde{d} \geq \delta$ ($\simeq$ 0.7 $\mu$m).
The first two conditions justify the derivation of Eq. (\ref{Eq:Vx})
and the third guarantees that the bulk insulating properties
are surely captured in the SMIS studies.
Regardless of the naive assumption that the SC state is suppressed over all 
the surfaces, this shell-type model well explains the present crystalline size effect.
Indeed, as shown in Fig. \ref{Fig:SMIS}(b), the SC region clearly 
remains in the \textit{Q} state.
We never rule out the existence of insulating regions that
penetrate further into the sample than $\tilde{d}$. 

The existence of a macroscopic insulating phase at a depth of a few $\mu$m
explains the overestimation of the fast cooling data for 
$\lambda_{\parallel}(0)$ observed in the low magnetic field measurements.\cite{785}
In the magnetization ($M$) measurements performed in magnetic fields higher than 
$H_{c1}$, $\lambda_{\parallel}$ is obtained from 
the slope in the $M(H)$ curves, in which
the influence of the insulating regions is sufficiently negligible.\cite{lambda}
However, in the low field measurements,
the correct estimation of $\lambda_{\parallel}$ is
disturbed by the shell-type insulating domain induced by fast cooling, because the 
magnetic field penetrates into the insulating regions.
This leads to the overestimation of $\lambda_{\parallel}(0)$ as 
described below.
If the suppression of the SC volume is attributed to the magnetic 
penetration depth to the SC state from the edge,
$\lambda_{\parallel}(0)$ becomes very large, $\sim$ 24--100 $\mu$m.\cite{785}
However, this is an incorrect treatment because the suppressed SC volume fraction 
after quenching mainly contains the contribution of the insulating shell 
with $\tilde{d}$, as demonstrated in Fig.\ref{Fig:Vx}. 
Accordingly, the very large $\lambda_{\parallel}(0)$ in the low field 
measurements after fast cooling can be explained as being due to the 
influence of the shell-type insulating domain.

Finally we discuss what happens after fast cooling in $\kappa$-h8ET-Br.
Two contributions originate from fast cooling.
The first is the chemical pressure effect, as in the 
case of deuteration. 
As discussed in previous reports,\cite{203,818} fast cooling 
through 80 K decreases $W$, giving rise to an increase of $U/W$.
Thus, the macroscopic non-SC region in quenched $\kappa$-h8ET-Br 
will also be a Mott insulating phase,
because $\kappa$-h8ET-Br is in the vicinity of the Mott insulating phase.
Indeed, no SC suppression is detected in $\kappa$-(h8ET)$_2$Cu(NCS)$_2$,
which is farther from the Mott boundary.
The present size effect on the SC volume and the large increase of 
$\lambda_{\parallel}(0)$ observed in the low field 
measurements\cite{785} are due to the existence of the macroscopic 
AFI region induced by chemical pressure.

The second contribution from fast cooling is the introduction of molecular disorder.
There is an internal degree of freedom in the terminal ethylene groups
of the ET molecule (conformational disorder).
The conformational disorder is thermally excited at room
temperature\cite{692} and is frozen into the system by fast cooling.
The frozen disorder will be distributed randomly all over 
the sample, both in the SC and non-SC regions.
The disorder in the SC region gives impurity scattering.
The increase of the Dingle temperature\cite{278} and the slight increase 
of $\lambda_{\parallel}(0)$ observed in the high field 
measurements\cite{220,1151} can be explained as the influence of 
quasi-particle scattering by the increase of disorder in the fast 
cooling process.
Meanwhile, a decrease of $T_{c}$,\cite{220,1151,203,146} vortex pinning suppression,\cite{1030} and 
an increase of the residual resistivity\cite{146} 
may result from both the contributions of microscopic impurity scattering and 
macroscopic insulating region (or boundary) formation by fast cooling.

The question as to why the insulating region grows in the vicinity of the surface is 
still open.
A possible speculation is that much more disorders in the surface 
vicinity are induced by very fast cooling than those in the inner site.
A very fast cooling treatment may not guarantee a homogeneous cooling state
within a sample, because of a thermal flow problem.
Such inhomogeneous distribution of disorders may result in the present
shell type insulating state.
Besides, several authors have reported unconventional electronic properties 
arising from the surface state.\cite{1157,965,310}
For example, in $\kappa$-(h8ET)$_2$Cu(NCS)$_2$, two kinds of photoemission
spectra have been found, attributed to the different surface 
types.\cite{1157}
However, the present shell-type insulating domain has a bulk scale of $\sim$ 2 $\mu$m 
in depth, which is about 700 times larger than the inter-layer
distance $\sim$ 15 \AA.
This seems to be too large as a coherence scale originating from the surface structure
of the first layer, such as structural reconstruction.
Further study will be needed to clarify this point.

\section{Conclusion}
In conclusion, we have proposed a macroscopic shell-type insulating domain 
in quenched $\kappa$-h8ET-Br by the observation of a marked size effect on the 
SC suppression and by SMIS mapping of the \textit{Q} state.
The thin insulating domain is located in the surface vicinity with 
50--100 $\mu$m diameter and a few $\mu$m depth.
The very large value of $\lambda_{\parallel}(0)$ obtained from the 
measurements in magnetic fields lower than $H_{c1}$ after fast cooling
can be explained well by the existence of the shell-type insulating domain.

\begin{acknowledgments}
The authors thank T.~Hirono, T.~Kawase, and T.~Moriwaki for their 
technical support.
We are also greatful to K. Miyagawa for helpful comments.
Synchrotron radiation measurements were performed at SPring-8 with the approval of JASRI 
(2003A0075-NS1-np and 2003B0114-NSb-np).
This research was supported by a Grant-in-Aid for Scientific Research (C) 
from JSPS.
\end{acknowledgments}


\begin{references}
\bibitem{size_Br}The sample dimensions in 
$\kappa$-(h8ET)$_2$Cu[N(CN)$_2$]Br for the static susceptibility 
measurements are as follows:
$x\times y \times z$ in \#1 is $1.84\times1.57\times0.34$,\cite{1030}
\#2: $1.09\times1.06\times0.13$,\cite{1151}
\#3: $1.04\times0.96\times0.25$,\cite{1210}
\#4: $0.84\times0.75\times0.21$,
\#5: $0.71\times0.67\times0.17$,
\#6: $0.55\times0.47\times0.13$,
\#7: $0.42\times0.39\times0.10$,
\#8: $0.40\times0.37\times0.07$,
\#9: $0.27\times0.24\times0.04$,
and \#10: $0.26\times0.24\times0.07$ mm$^3$.
\bibitem{size_CuNCS}The sample dimensions in 
$\kappa$-(h8ET)$_2$Cu(NCS)$_2$ for the static susceptibility 
measurements are as follows;
$x\times y \times z$ in \#1 is $2.2\times1.1\times0.40$,\cite{1151}
\#2: $2.5\times1.4\times0.26$,
\#3: $0.72\times0.41\times0.08$,
and \#4: $0.32\times0.20\times0.07$ mm$^3$.
\bibitem{lambda}
Provided all the crystalline volume of $\kappa$-(h8ET)$_2$Cu[N(CN)$_2$]Br 
in the sample of Ref. \onlinecite{1151} become the AFI state, the 
maximum value of $M$ is approximately 2$\times 10^{-3}$ emu,  
with assuming $S$=1/2 per ET dimer.
This is two orders of magnitude lower than the reversible SC 
magnetization, $M \geq 0.1$ emu (see Fig. 6 in Ref. \onlinecite{1151}),
indicating the negligible contribution of the insulating thin domain
to the magnetization measurements in the mixed state.
\end{references}
\end{document}